\documentclass[prl,twocolumn,superscriptaddress,amssymb]{revtex4}
\usepackage{graphicx}
\usepackage{bm}
\usepackage{amsmath,amssymb}
\usepackage{color}
\usepackage{amsthm} 

\begin{document}
\title{Verified measurement-based quantum computing with hypergraph states} 
\author{Tomoyuki Morimae}
\email{morimae@gunma-u.ac.jp}
\affiliation{ASRLD Unit,
Gunma University, 1-5-1 Tenjincho Kiryushi
Gunma, 376-0052, Japan}
\author{Yuki Takeuchi}
\email{takeuchi@qi.mp.es.osaka-u.ac.jp}
\affiliation{Graduate School of Engineering Science,
Osaka University, Toyonaka, Osaka 560-8531, Japan}
\author{Masahito Hayashi}
\email{masahito@math.nagoya-u.ac.jp}
\affiliation{Graduate School of Mathematics, Nagoya University,
Furocho, Chikusaku, Nagoya, 464-8602, Japan}
\affiliation{Centre for Quantum Technologies, National University
of Singapore, 3 Science Drive 2, 117542, Singapore}

\begin{abstract}
Hypergraph states are generalizations of graph states where
controlled-$Z$ gates on edges
are replaced with generalized controlled-$Z$ gates on hyperedges.
Hypergraph states have several advantages
over graph states. For example,
certain hypergraph states, such as the Union Jack states,
are universal resource states for measurement-based quantum 
computing with only Pauli measurements,
while graph state measurement-based quantum computing
needs non-Clifford basis measurements.
Furthermore, it is impossible to classically efficiently sample
measurement results on hypergraph states 
with a constant $L$1-norm error
unless the polynomial hierarchy collapses to the third level.
Although several protocols 
have been proposed to verify graph states with
only sequential single-qubit Pauli measurements,
there was no verification method for hypergraph states.
In this paper, we propose a method for verifying
hypergraph states with only sequential
single-qubit Pauli measurements.
As applications, we consider verified 
blind quantum computing with hypergraph states,
and
quantum supremacy demonstrations with
hypergraph states.
\end{abstract}
\maketitle

Many-point correlations in quantum many-body systems
are one of the most essential ingredients 
in condensed-matter physics and statistical physics.
Correlations of sequential single-qubit measurements on quantum states
are also important drive forces for quantum information
processing.
For example,
measurement-based quantum computing~\cite{MBQC}, which is nowadays
one of the standard quantum computing models,
enables universal quantum computing with only adaptive single-qubit
measurements on certain quantum states,
such as graph states~\cite{MBQC} 
and other condensed-matter-physically motivated
states including the AKLT state~\cite{BrennenMiyake,MBQC1,MBQC2,
MBQC3,MBQC4,MBQC5,MBQC6,MBQC7,MBQC8,MBQC9,MBQC10,MBQC11,
MBQC12,MBQC13,MBQC14,MBQC15}. 
Furthermore,
not only adaptive but also non-adaptive single-qubit measurements
on graph states can demonstrate a quantumness which cannot be classically
efficiently simulated: 
it is known that
if probability distributions of
non-adaptive sequential single-qubit measurements
on graph states are classically efficiently sampled,
then the polynomial hierarchy collapses to the 
third level~\cite{IQP1,IQP2,FujiiMorimae}
or the second level~\cite{KobayashiICALP}.
The polynomial hierarchy is a hierarchy of complexity classes
generalizing P and NP,
and it is not believed to collapse
in computer science. 
It is an example of recently well studied
``quantum supremacies" of sub-universal quantum computing models, 
which are expected to be easier to experimentally implement,
but can outperform classical computing. (For details, see 
Refs.~\cite{bosonsampling,bosonsampling2,
MFF,KobayashiICALP,FujiiMorimae,IQP1,IQP2}
and their supplementary materials.)

For practical implementations of measurement-based quantum computing
and experimental demonstrations of the quantum supremacy,
verifying graph states is essential,
since in reality 
a generated state cannot be the ideal graph state
due to some experimental noises.
The problem becomes more serious if we consider
delegated secure quantum computing, so called blind quantum
computing~\cite{BFK,FK}.
It is known that the ability of sequentially
measuring single qubits is enough to secretly delegate quantum computing
to a remote server~\cite{measuringAlice,HayashiMorimae}. 
The honest server sends each qubit of a graph state one by one
to the user, and user can realize any quantum computing with
only sequential single-qubit measurements.
If the server is malicious, however,
a completely wrong state might be sent to the user.
The user therefore needs to test the state
sent from the server.
In such a quantum cryptographic scenario, the situation is worse
than the single-party laboratory experiments,
since the noises on the given state are caused by malicious servers
and therefore not necessarily physically natural ones.
Several methods of verifying graph states
with only sequential single-qubit Pauli measurements
have been proposed~\cite{HayashiMorimae,MorimaeNagajSchuch}.
(If more than two non-communicating servers are available,
a completely classical user can verify stabilizer 
states~\cite{VaziraniNature,Matt,Ji}.)
In the protocol of Ref.~\cite{HayashiMorimae}, 
the user does a test so called the stabilizer test
on some parts of the state sent from the server.
The stabilizer test can be done with only sequential
single-qubit Pauli measurements.
If the user passes the test, the remaining state is guaranteed
to be close to the ideal graph state.

Since the protocol of Ref.~\cite{HayashiMorimae}
makes no assumption (such as the i.i.d. sample or physically natural
noises) on the given state, the verification
method can be used in quantum cryptographic contexts.
In particular, verified blind quantum computing
and verified quantum supremacy demonstrations 
can be realized with graph states verified through
the protocol.
There are, however, two problems.
First, in the verified blind protocol of Ref.~\cite{HayashiMorimae},
the user needs non-Clifford basis measurements for computing
(the verification itself can be done with only Pauli measurements).
It would be better if both the
verification and the computation can be done with
only Pauli measurements~\cite{LH}.
Second, the quantum supremacy demonstration with
graph states~\cite{IQP1}, which needs only non-adaptive measurements,
requires somehow a strict approximation,
namely a multiplicative-error approximation.

Recently, two breakthroughs that solve these drawbacks
of graph states have been done.
These results use
hypergraph states~\cite{Rossi,HG1,HG2,HG3,HG4} in stead of graph states.
(For the definition of hypergraph states and their
properties, see below.)
First, 
certain hypergraph states, such as the Union Jack states,
are universal resource states for measurement-based quantum 
computing with only Pauli measurements~\cite{MillerMiyake}.
This result solves the first problem,
namely, the requirement of non-Clifford basis
measurements for the user.
Therefore, by using the hypergraph states, the one-way
secure delegated quantum computing is possible for the
user who can do only Pauli measurements.
Ref.~\cite{MillerMiyake} also pointed out that
hypergraph states are important in the study
of symmetry-protected topological orders.
Second, it was shown in Ref.~\cite{IQP2} that
if hypergraph states are considered,
the multiplicative error requirement
can be replaced with an $L$1-norm one,
which is more relaxed.
This result solves the second problem. 

In short, hypergraph states are promising novel resource states for
many quantum information processing tasks.
Unfortunately, however, there was no verification method for hypergraph states.
In particular, we did not know how to test 
a given hypergraph state with only sequential single-qubit
Pauli measurements.
It was a huge obstacle for practical applications of hypergraph states
in quantum information and condensed matter physics.

In this paper, we propose a method for verifying
hypergraph states with only sequential
single-qubit Pauli measurements.
As in the case of the graph state 
verification~\cite{HayashiMorimae}, the user
does a certain test on some parts of the state sent from the server.
If the user passes the test, then the remaining state is guaranteed to be
close to the ideal hypergraph state.
As applications, we consider
verified blind quantum computing with hypergraph states,
and verified quantum supremacy demonstrations with
hypergraph states.

{\it Hypergraph states}.---
We first define hypergraph states, and explain their
properties.
A hypergraph $G\equiv(V,E)$ 
is a pair of a set $V$ of vertices and a set $E$ of hyperedges,
where $n\equiv |V|$.
A hyperedge may link more than two vertices.
For simplicity, in this paper, we assume that
$2\le|e|\le3$ for all $e\in E$, 
where $|e|$ is the number of vertices linked to the
hyperedge $e$.
(Generalizations to other cases would be possible.)
Let 
\begin{eqnarray*}
|G\rangle\equiv \Big(\prod_{e\in E}\widetilde{CZ}_e\Big)|+\rangle^{\otimes n}
\end{eqnarray*}
be the hypergraph state corresponding to the hypergraph $G$,
where 
$
\widetilde{CZ}_e\equiv 
\bigotimes_{i\in e}I_i-2\bigotimes_{i\in e}|1\rangle\langle 1|_i
$
is the generalized $CZ$ gate acting
on vertices in the hyperedge $e$.
Here, $I$
is the two-dimensional identity operator.
For example, if $|e|=2$, it is nothing but the standard $CZ$ gate.
If $|e|=3$, it is the $CCZ$ gate,
$
CCZ\equiv
(I^{\otimes 2}-|11\rangle\langle11|)\otimes I
+|11\rangle\langle11|\otimes Z.
$

The stabilizer $g_i$ of $|G\rangle$ associated with the vertex $i$ is
defined by
\begin{eqnarray*}
g_i&\equiv&
\Big(\prod_{e\in E}\widetilde{CZ}_e\Big)
X_i
\Big(\prod_{e\in E}\widetilde{CZ}_e\Big)\\
&=&X_i
\Big(\prod_{j\in W_i^Z}Z_j\Big)
\Big(\prod_{(j,k)\in W_i^{CZ}}CZ_{j,k}\Big),
\end{eqnarray*}
where
\begin{eqnarray*}
W_i^Z&\equiv&\{j\in V~|~(i,j)\in E\},\\
W_i^{CZ}&\equiv&\{(j,k)\in V\times V~|~(i,j,k)\in E\}.
\end{eqnarray*}
It is easy to check
that the following properties are satisfied:
$[g_i,g_j]=0$ for all $i,j\in V$.
$g_i|G\rangle=|G\rangle$
for all $i\in V$. 
$g_i^2=I^{\otimes n}$ for all $i\in V$. 
$
\prod_{i=1}^n\frac{I^{\otimes n}+g_i}{2}
=|G\rangle\langle G|.
$

{\it Stabilizer test for $g_i$}.---
Before introducing our verification protocol,
we define the stabilizer test for each $g_i$, which is an
essential ingredient of the protocol.
Note that
$
CZ_{j,k}=\frac{1}{2}(I_j\otimes I_k+
I_j\otimes Z_k+Z_j\otimes I_k-Z_j\otimes Z_k).
$
Therefore
\begin{eqnarray*}
g_i&=&X_i\Big(\prod_{j\in W_i^Z}Z_j\Big)
\Big(
\frac{1}{2^r}\sum_{t\in\{1,2,3,4\}^r}
\prod_{(j,k)\in W_i^{CZ}}\sigma_{j,k}(t_{j,k})
\Big)\\
&=&\frac{1}{2^r}
\sum_{t\in\{1,2,3,4\}^r}
s_t,
\end{eqnarray*}
where $r\equiv|W_i^{CZ}|$,
$t\equiv\{t_{j,k}\}_{(j,k)\in W_i^{CZ}}$,
$\sigma_{j,k}(1)\equiv I_j\otimes I_k$,
$\sigma_{j,k}(2)\equiv I_j\otimes Z_k$,
$\sigma_{j,k}(3)\equiv Z_j\otimes I_k$,
$\sigma_{j,k}(4)\equiv -Z_j\otimes Z_k$,
and
\begin{eqnarray*}
s_t
&\equiv&
X_i\Big(\prod_{j\in W_i^Z}Z_j\Big)
\Big(
\prod_{(j,k)\in W_i^{CZ}}\sigma_{j,k}(t_{j,k})
\Big).
\end{eqnarray*}
Let us define a bit $\alpha_t\in\{0,1\}$
and a subset $D_t\subseteq V$ such that
\begin{eqnarray*}
s_t
=
(-1)^{\alpha_t}X_i\Big(\prod_{j\in D_t}Z_j\Big).
\end{eqnarray*}
Note that $\alpha_t$ and $D_t$ can be calculated
in polynomial time.
In fact, $\alpha_t$ can be calculated in the following way.
We first set $\alpha_t=0$.
If $t_{j,k}=4$, we flip $\alpha_t$.
We do it for all $(j,k)\in W_i^{CZ}$.
Since $|W_i^{CZ}|\le {n-1 \choose 2}=O(n^2)$,
it takes at most polynomial time.
Furthermore, $D_t$ can be calculated in the following way.
We first set $D_t=W_i^Z$.
We then update $D_t$ according to $t_{j,k}$ for each $(j,k)\in W_i^{CZ}$.
Again, $|W_i^{CZ}|\le O(n^2)$ means that it takes
at most polynomial time.

Let $\rho$ be an $n$-qubit state.
We define the ``stabilizer test for $g_i$ on $\rho$" as the following
Alice's action:
\begin{itemize}
\item[1.]
Alice randomly generates $t\in\{1,2,3,4\}^r$.
\item[2.]
She measures $i$th vertex of $\rho$ in $X$, and
$j$th vertex of $\rho$ in $Z$ for all $j\in D_t$. 
\end{itemize}
Let $x\in\{+1,-1\}$ be the
measurement result of the $X$ measurement,
and $z_j\in\{+1,-1\}$ be that of the $Z$ measurement
on vertex $j\in D_t$.
We say that Alice passes the stabilizer test for $g_i$ on $\rho$ if 
$
x\prod_{j\in D_t}z_j=(-1)^{\alpha_t}.
$

The probability $p_{{\rm test},i}$ that Alice passes
the stabilizer test for $g_i$ on $\rho$ is~\cite{expr}
\begin{eqnarray*}
p_{{\rm test},i}
\equiv\frac{1}{4^r}\sum_{t\in\{1,2,3,4\}^r}
\mbox{Tr}\Big(\rho\frac{I^{\otimes n}+s_t}{2}\Big)
=\frac{1}{2}+
\frac{\mbox{Tr}(\rho g_i)}
{2^{r+1}}.
\end{eqnarray*}

\begin{figure}[htbp]
\begin{center}
\includegraphics[width=0.3\textwidth]{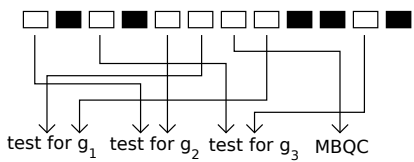}
\end{center}
\caption{
An example for $n=3$, $k=2$, $m=5$.
Each square represents a register that stores
$n$ qubits. Registers represented by black squares are discarded.
} 
\label{deFinetti}
\end{figure}

{\it Verification protocol}.---
We now explain our verification protocol.
Bob sends Alice an $n(nk+1+m)$-qubit 
state $\Psi$, 
where $k=2^{2r+3}n^7$ and $m\ge2n^7k^2\ln2$.
The state $\Psi$ consists of
$nk+1+m$ registers (Fig.~\ref{deFinetti}). 
Each register stores $n$ qubits.
(If Bob is honest, every register is in the state $|G\rangle$.
If Bob is malicious, on the other hand, $\Psi$ can be any
$n(nk+1+m)$-qubit entangled state.)
Alice randomly permutes registers and
discards $m$ registers.
(As we will see later,
this random permutation and discarding of some registers are
necessary to guarantee that the remaining state is close to
an i.i.d. sample by using the quantum de Finetti theorem~\cite{LiSmith}.)
Let $\Psi'$ be the remaining state.
The state $\Psi'$ consists of $nk+1$ registers.
She chooses one register from $\Psi'$, which is used for
the measurement-based quantum computing.
We call the register computing register.
The remaining $nk$ registers of $\Psi'$ are divided into
$n$ groups. Each group consists of $k$ registers.
The stabilizer test for $g_i$ is performed
on every register in the $i$th group for $i=1,2,...,n$.
(Note that Alice does not need to do the permutation ``physically",
which requires a quantum memory.
Bob just sends each qubit of $\Psi$ one by one to Alice,
and Alice randomly chooses her action from the test, discarding,
or computation.)

Let $K_i$ be the number of times that Alice
passes the stabilizer test for $g_i$, i.e. the random
variable to describe the number of Alice's observation
of the event $\frac{1}{4^r}\sum_t\frac{I^{\otimes n}+s_t}{2}$.
If 
$
\frac{K_i}{k}\ge\frac{1}{2}+\frac{1-\epsilon}{2^{r+1}},
$
we say that the $i$th group passes the test.
Here, $\epsilon=\frac{1}{2n^3}$.
If all groups pass the test, we say that 
Alice accepts Bob.

The main results of the present paper
are the following two items:

1. Completeness:
if every register of $\Psi$ is in the state
$|G\rangle$,
then the probability that Alice accepts Bob is
larger than
$1-ne^{-n}$.

2. Soundness:
if Alice accepts Bob, the state $\rho_{\rm comp}$
of the computing register
satisfies
$
\langle G|\rho_{\rm comp}|G\rangle\ge1-\frac{1}{n}
$
with a probability larger than $1-\frac{1}{n}$.

{\it Proof of the completeness}.---
We first show the completeness.
If every register of
$\Psi$ is in the state $|G\rangle$,
then
$
p_{{\rm test},i}=\frac{1}{2}+\frac{1}{2^{r+1}}
$
for all $i=1,2,...,n$.
From the union bound and the Hoeffding inequality,
\begin{eqnarray*}
\mbox{Pr}[\mbox{Alice accepts Bob}]
&=&\mbox{Pr}\Big[\bigwedge_{i=1}^n
\Big(\frac{K_i}{k}\ge\frac{1}{2}+\frac{1-\epsilon}{2^{r+1}}\Big)\Big]\\
&\ge&
1-\sum_{i=1}^n
\mbox{Pr}\Big[\frac{K_i}{k}< \frac{1}{2}+\frac{1-\epsilon}{2^{r+1}}\Big]\\
&=&
1-\sum_{i=1}^n
\mbox{Pr}\Big[\frac{K_i}{k}< p_{{\rm test},i}-\frac{\epsilon}{2^{r+1}}\Big]\\
&\ge&
1-n e^{-2\frac{\epsilon^2}{2^{2r+2}}k}.
\end{eqnarray*}

{\it Proof of the soundness}.---
We next show the soundness.
We define the $n$-qubit projection operator
$
\Pi_G^\perp\equiv I^{\otimes n}-|G\rangle\langle G|.
$
Let $T$ be the POVM element corresponding to
the event that Alice accepts Bob.
We can show that
for any $n$-qubit state $\rho$,
\begin{eqnarray}
\mbox{Tr}\Big[
(
T\otimes \Pi_G^\perp)
\rho^{\otimes nk+1}\Big]
\le
\frac{1}{2n^2}.
\label{eq1}
\end{eqnarray}
Its proof is given later.
Due to the quantum de Finetti theorem (for the one-way LOCC
norm version)~\cite{LiSmith},
\begin{eqnarray*}
\mbox{Tr}\Big[
(
T\otimes \Pi_G^\perp)
\Psi'\Big]
&\le&
\mbox{Tr}\Big[
(
T\otimes \Pi_G^\perp)
\int d\mu(\rho)\rho^{\otimes nk+1}\Big]\\
&&+\frac{1}{2}\sqrt{\frac{2n^2k^2n\ln 2}{m}}\\
&\le&\frac{1}{2n^2}+\frac{1}{2}\sqrt{\frac{2n^3k^2\ln2}{2n^7k^2\ln2}}
=\frac{1}{n^2}.
\end{eqnarray*}
We have
$
\mbox{Tr}[
(T\otimes\Pi_G^\perp)\Psi']
=
\mbox{Tr}(\Pi_G^\perp\rho_{\rm comp})
\mbox{Tr}[(T\otimes I)\Psi'].
$
Therefore,
if $\mbox{Tr}(\Pi_G^\perp\rho_{\rm comp})>\frac{1}{n}$,
then
$
\mbox{Tr}[(T\otimes I)
\Psi']
<
\frac{1}{n},
$
which means that if Alice accepts Bob,
$
\langle G|\rho_{\rm comp}|G\rangle\ge1-\frac{1}{n}
$
with a probability larger than $1-\frac{1}{n}$.

{\it Proof of Eq.~(\ref{eq1})}.---
First, let us assume
that
$\mbox{Tr}(\rho g_i)\ge 1-\delta$ for all $i=1,2,..,n$,
where $\delta=\frac{1}{n^3}$. 
Due to the union bound,
\begin{eqnarray*}
1-\langle G|\rho|G\rangle&=&
1-\mbox{Tr}\Big(\prod_{i=1}^n\frac{I^{\otimes n}+g_i}{2}\rho\Big)\\
&\le&\sum_{i=1}^n\Big[
1-\mbox{Tr}\Big(\rho\frac{I^{\otimes n}+g_i}{2}\Big)\Big]
\le\frac{n\delta}{2}.
\end{eqnarray*}
Therefore, 
\begin{eqnarray}
\mbox{Tr}\Big[
(T\otimes \Pi_G^\perp)
\rho^{\otimes nk+1}\Big]
&=&
\mbox{Tr}(T\rho^{\otimes nk})
\mbox{Tr}(\Pi_G^\perp\rho)\nonumber\\
&\le& 1\times \frac{n\delta}{2}
=\frac{1}{2n^2}.
\label{firstcase}
\end{eqnarray}

Next let us assume that
$\mbox{Tr}(\rho g_i)< 1-\delta$ for at least one $i$.
In this case,
\begin{eqnarray*}
p_{{\rm test},i}=\frac{1}{2}+\frac{\mbox{Tr}(\rho g_i)}{2^{r+1}}
<\frac{1}{2}+\frac{1-\delta}{2^{r+1}}
\end{eqnarray*}
for the $i$.
Then, due to the Hoeffding inequality,
\begin{eqnarray*}
\mbox{Tr}[(T\otimes I)\rho^{\otimes nk+1}]
&\le& \mbox{Pr}[\mbox{group $i$ passes the test}]\\
&=&\mbox{Pr}\Big[\frac{K_i}{k}\ge\frac{1}{2}+\frac{1-\epsilon}{2^{r+1}}\Big]\\
&=&\mbox{Pr}\Big[
\frac{K_i}{k}\ge\frac{1}{2}+\frac{1-\delta}{2^{r+1}}
+\frac{\delta-\epsilon}{2^{r+1}}\Big]\\
&\le&\mbox{Pr}\Big[
\frac{K_i}{k}>p_{{\rm test},i}+\frac{\delta-\epsilon}{2^{r+1}}\Big]\\
&\le& e^{-2\frac{(\delta-\epsilon)^2}{2^{2r+2}}k}
= e^{-n}.
\end{eqnarray*}
Hence
\begin{eqnarray}
\mbox{Tr}[(T\otimes\Pi_G^\perp)\rho^{\otimes nk+1}]
&=&
\mbox{Tr}(T\rho^{\otimes nk})
\mbox{Tr}(\Pi_G^\perp\rho)\nonumber\\
&\le&
e^{-n}\times 1.
\label{secondcase}
\end{eqnarray}

From Eqs.~(\ref{firstcase}) and (\ref{secondcase}),
for any state $\rho$,
\begin{eqnarray*}
\mbox{Tr}[(T\otimes\Pi_G^\perp)\rho^{\otimes nk+1}]
\le\max\Big(\frac{1}{2n^2},e^{-n}\Big)
=\frac{1}{2n^2}.
\end{eqnarray*}

{\it Applications}.---
To conclude this paper,
we finally discuss two applications of our results.
First, our verification protocol can be used in 
verified blind quantum computing. In the protocol of 
Ref.~\cite{HayashiMorimae}, the user needs non-Clifford basis measurements
to implement quantum computing (the verification itself can be done
with only Pauli measurements.)
On the other hand, if the server generates
the Union Jack states~\cite{MillerMiyake}, 
for example, the user needs only Pauli measurements for both 
the verification and the computation.

Second, our protocol can be used for the verified quantum supremacy
demonstration. It was shown in Ref.~\cite{IQP2} that 
the following is true for several hypergraph states
(assuming the so called ``worst case vs average case" conjecture):
if there exists a classical sampler that outputs $z$
with probability $q_z$ such that
$
\sum_{z\in\{0,1\}^n}|p_z-q_z|\le\frac{1}{192},
$
then the polynomial hierarchy collapses to the third level.
Here, $p_z$ is the probability of obtaining
the result $z\in\{0,1\}^n$ when certain single-qubit measurements are
done on an $n$-qubit hypergraph state.
This result means that
if we can generate hypergraph states, we can demonstrate
the quantum supremacy. However, what happens if we cannot
have the ideal hypergraph state, and only
the verified state $\rho_{\rm comp}$ is available?
(For example, Alice, who can do only single-qubit measurements,
might want untrusted Bob to
send a hypergraph state.)
We can show that the state $\rho_{\rm comp}$ is enough to demonstrate
the same quantum supremacy. In fact,
let us assume that there exists a classical sampler
such that
$
\sum_z|p_z'-q_z|\le\frac{1}{192},
$
where
$p'_z$ is the output probability distribution
of the single-qubit measurements on $\rho_{\rm comp}$.
Then,
\begin{eqnarray*}
\sum_z|p_z-q_z|&\le&
\sum_z|p_z-p_z'|+\sum_z|p_z'-q_z|\\
&\le& 
o(1)
+\frac{1}{192},
\end{eqnarray*}
which means that the classical sampler can also sample
$p_z$ with the $\sim1/192$ $L$1-norm error.
For example, the hypergraph states
naturally induced from
the IQP circuits corresponding to
the non-adaptive
Union Jack state measurement-based quantum computing~\cite{MillerMiyake}
can be used for that purpose.
Since the non-adaptive Union Jack state measurement-based
quantum computing is universal
with postselections,
a multiplicative error calculation
of its output probability distribution
is $\#$P-hard~\cite{FujiiMorimae}. 
If we assume the worst case hardness can be
lifted to the average case one,
we can show the hardness of 
the classical constant $L$1-norm error sampling.

TM is supported by the JST ACT-I, 
the JSPS Grant-in-Aid for Young Scientists (B) No.26730003, and 
the MEXT JSPS Grant-in-Aid for Scientific Research on Innovative Areas No.15H00850.
YT is supported by the Program for Leading Graduate Schools: Interactive
Materials Science Cadet Program.
MH is supported in part by Fund for the Promotion of Joint
International Research (Fostering Joint International Research)
No.15KK0007, the JSPS MEXT Grant-in-Aid for Scientific
Research (B) No.16KT0017, the Okawa Research Grant and Kayamori
Foundation of Information Science Advancement.


\begin{thebibliography}{00}
\bibitem{LH}
If a BQP problem is mapped to a local Hamiltonian problem,
Pauli measurements are enough to do the verified computation.
However, there are two problems for the idea.
First, the server needs to generate somehow complicated
states, so called Kitaev-Feynmann history states.
Second, it is no longer blind, since the server has to know
the program and input to generate Kitaev-Feynmann history states.
Another idea would be to embed magic states in the graph state
in advance.
Pauli measurements are enough for universal quantum computing on
such ``magic-state-embedded" graph states.
However, the verification of them seems to be more complicated,
since we have to verify both the graph state
and magic states at the same time.
\bibitem{expr}
If $r=poly$, then $p_{{\rm test},i}=\frac{1}{2}+O(2^{-poly})$,
which means that exponentially many measurements are required to gain
useful information about $\mbox{Tr}(\rho g_i)$.
Therefore, our verification method cannot be used
if $r=poly$.
For many applications, however, $r=const.$, such as
the Union Jack states.
It is an open problem whether we can verify
hypergraph states with $r=poly$.
\end{thebibliography}
\end{document}